\documentstyle[preprint,revtex]{aps}

\begin{document}
\draft
\begin{title}
Theory of Thermodynamic Magnetic Oscillations in Quasi-One-Dimensional
Conductors
\end{title}
\author{Victor M. Yakovenko\cite{*}}
\begin{instit}
Serin Physics Laboratory, Rutgers University, P.O.Box 849, Piscataway, NJ 08855
\end{instit}
\receipt{1992}
\begin{abstract}

   The second order correction to free energy due to the interaction
between electrons is calculated for a quasi-one-dimensional conductor
exposed to a magnetic field perpendicular to the chains.  It is found
that specific heat, magnetization and torque oscillate when the magnetic
field is rotated in the plane perpendicular to the chains or when the
magnitude of magnetic filed is changed.  This new mechanism of
thermodynamic magnetic oscillations in metals, which is not related to
the presence of any closed electron orbits, is applied to explain
behavior of the organic conductor (TMTSF)$_2$ClO$_4$.

\end{abstract}

\pacs{PACS numbers: 71.25.Hc, 71.10.+x, 74.70.Kn, 75.20.En}

\narrowtext

   In a recent study \cite{N} of the quasi-one-dimensional conductor
(TMTSF)$_2$ClO$_4$, a new physical effect has been found.  In this
experiment, a magnetic field was rotated in the plane perpendicular to
the conducting chains and the torque along the chains was measured.  The
torque, as a function of the angle $\theta$ between the magnetic field
and the $c$-axis, was observed to exhibit oscillations approximately
periodic in $\tan\theta$.  Since the torque is a derivative of the free
energy $F$ with respect to $\theta$, then $F$ must also oscillate upon
rotation of the magnetic field.  In earlier experiments, magnetization
\cite{YC} and specific heat \cite{F} were measured to be oscillating
functions of the magnetic field magnitude (the so called ``fast''
oscillations).  Such angular and fast thermodynamic magnetic
oscillations were totally unexpected theoretically for a
quasi-one-dimensional metal.  In momentum space, an electron in a
magnetic field moves along an orbit obtained as the intersection of a
constant energy surface and a plane perpendicular to the magnetic field.
Magnetic oscillations of the free energy (de Haas-van Alfven effect) are
known in the situation when the orbits at the Fermi surface are closed.
In this case, the orbits are quantized, the energy spectrum is discrete
and the oscillations in $F$ are due to the Landau levels crossing the
Fermi level.  However, in a quasi-one-dimensional conductor, the Fermi
surface consists of two slightly warped planes perpendicular to the
direction of the chains.  Thus, if a magnetic field has been applied
perpendicular to the chains, all electron orbits at the Fermi surface
are open, the energy spectrum is continuous and no oscillations in $F$
are expected for non-interacting electrons \cite{LAK}.  In the present
paper, the lowest order correction to $F$ due to the interaction between
electrons is calculated for a quasi-one-dimensional conductor and is
shown to exhibit angular and fast magnetic oscillations.

   The phenomenon of angular oscillations in a quasi-one-dimensional
conductor was predicted theoretically by Lebed \cite{L} who calculated
the oscillations of the magnetic-field-induced spin-density-wave
(FISDW) transition temperature.  Experiment \cite{XY} did not reveal
these oscillations, however, the studied range of angles might not
have been big enough to see the first oscillation \cite{N}.  The
angular oscillations in the conductivity were predicted \cite{LB} and
observed \cite{MRS,Oe,N}.  Two other models for the oscillations in
the conductivity were later suggested \cite{Ot}. None of mentioned
theoretical papers considered the oscillations of free energy.
Recently, the contribution of the FISDW order parameter fluctuations
to $F$ has been calculated and been shown to exhibit angular
oscillations \cite{M2}.  In the present paper, the approach of Ref.
\cite{L} and \cite{LB} is used to calculate the lowest order
correction to $F$.  This theory is applicable in the metallic phase
sufficiently far from the phase boundary, while theory \cite{M2}
is applicable in the vicinity of the phase transition to the FISDW
state.

   Let us consider a square lattice of parallel chains with a spacing
$b$ between the chains, $x$-axis parallel the chains.
The electron hopping
integrals between the nearest chains in the $y$ and $z$ directions are
equal to $t_b$ and $t_c$, respectively; $t_c\ll t_b\ll E_F$,
where $E_F$ is the Fermi energy.  Magnetic field $H$ is
applied in the $(y,z)$ plane at an angle $\theta$ measured from
$z$-axis. The electron Hamiltonian without the interaction term can be
written as \cite{GL}:
\begin{eqnarray}
&&\hat{H}_{\alpha\sigma}=-\alpha iv_F\partial_x+2t_b\cos(k_y-ebHx\cos\theta/c)
\nonumber\\
&&+2t_c\cos(k_z-ebHx\sin\theta/c)-\mu_BH\sigma,
\label{H}
\end{eqnarray}
where the index $\alpha=\pm$ labels electrons whose momenta along the
chains are close to $\pm k_F$, $v_F$ and $k_F$ are the Fermi velocity
and the Fermi momentum, $k_y$ and $k_z$ are momenta along the $y$- and
$z$-axes, $\sigma$ is the spin index, $\mu_B$ is the Bohr magneton,
$e$ is the electron charge, $c$ is the velocity of light and
$\hbar\equiv1$.  The two cosine terms in (\ref{H}), which describe
electron motion perpendicular to the chains, become in magnetic
field two periodic potentials along the chains.  Their periods are
commensurate at rational values of $\tan\theta$. This effect is
responsible for the angular oscillations of free energy.

   The lowest order non-trivial correction to the free energy $\Delta
F$ due to the interaction between electrons is given by the diagram
shown in Fig. 1.  Using Green functions, derived from Eq. (\ref{H})
\cite{GL}, one can find the following expression for the free energy
per electron:
\begin{eqnarray}
&&\Delta F=-\frac{2\pi^3g^2T^3}{\varepsilon_F\Omega}\int_0^\infty dx
\frac{\cosh(\tau x)}{\sinh^3(\tau x)}f_y(x)f_z(x),\label{DF} \\
&&f_y(x)=\langle J_0^2(V_b\sin(x\cos\theta)\sin k/\cos\theta)\rangle_k,
\label{fy} \\
&&f_z(x)=\langle J_0^2(V_c\sin(x\sin\theta)\sin k/\sin\theta)\rangle_k.
\label{fz}
\end{eqnarray}
Here $\varepsilon_F\equiv v_Fk_F$, $T$ is the temperature,
$\Omega=ebv_FH/c$ is the characteristic energy of magnetic field,
$\tau=4\pi T/\Omega$, $V_{b(c)}=8t_{b(c)}/\Omega$, $J_0$ is the Bessel
function and $\langle...\rangle_k$ means the averaging over $k$.
$g^2=(g_2-g_1)^2+g_2^2$, where $g_1$ and $g_2$ are dimensionless
interaction constants between electrons, normalized in such a way that
the mean-field SDW transition temperature is proportional to
$\exp(-1/g_2)$. Another contribution to $\Delta F$, which comes from
spin reversal scattering, contains an additional factor
$\cos(4\mu_BHx/\Omega)$ under the integral in Eq. (\ref{DF}) and
$g_1^2$ appears instead of $g^2$.  This contribution is more complex
and will not be studied here.

   Integral (\ref{DF}) is divergent at small $x$.  However, observable
quantities considered below are given by convergent integrals.
The correction to the specific heat per electron,
derived from Eq. (\ref{DF}), takes the form: \FL
\begin{equation}
\Delta C=-\frac{3}{4}g^2C_0\int_0^\infty dx
\left(\frac{x^2}{\sinh^2x}\right)'''
f_y(\frac{x}{\tau})f_z(\frac{x}{\tau}),
\label{DC}
\end{equation}
where the primes denote the third derivative with respect to $x$ and
$C_0=\pi^2T/3\varepsilon_F$ is the specific heat of non-interacting
electrons.  In Fig. 2, $\Delta C(\theta)$ is shown for three different
values of $\tau$ (0.1, 0.2 and 0.4) at $V_b=20$ and $V_c=2$. $\Delta C$
exhibits sharp peaks when $\tan\theta$ attains rational values, which is
where the periods of the functions $f_y(x)$ and $f_z(x)$ in (\ref{DC})
become commensurate.  The widths of the peaks go to zero when
$T\rightarrow0$ and the curve becomes fractal.  Each peak is split
because the term $(x^2/\sinh^2x)'''$ in (\ref{DC}) oscillates once over
the range $[0,\infty]$.  This splitting vanishes as $T\rightarrow0$.  It
follows from (\ref{fy}) and (\ref{fz}) that the transformation
$\theta\rightarrow45^\circ-\theta$ interchanges $V_b$ and $V_c$ in
(\ref{DC}), having small effect on the plots $\Delta C(\theta)$ in Fig.
2, despite the employed values of $V_b$ and $V_c$ differ by an order of
magnitude.  This means that $\Delta C$ does not depend qualitatively on
the precise values of $V_b$ and $V_c$ if they are greater than 2.
According to Ref. \cite{Y2}, in (TMTSF)$_2$ClO$_4$, $\Omega/H=1.8$ K/T,
but due to the unit cell doubling the two times bigger value should be
used.  Thus, the field e.g. 5 T corresponds to $\Omega=18$ K and the
values of the parameters $\tau$ and $V$ in Fig. 2 correspond to
$T=0.14$K, 0.29K, 0.57K, $t_b=45$K and $t_c=4.5$K.  The dependence of
specific heat on the orientation of the magnetic field has not been
measured yet experimentally.

   The correction to the magnetization per electron along the $z$-axis
$\Delta M_z=-\partial\Delta F/\partial H_z$, where $H_z=H\cos\theta$, is
given by the following expression:\FL
\begin{eqnarray}
&&\Delta M_z=\frac{g^2t_b\Omega}{2\varepsilon_FH}\int_0^\infty dx
\frac{\cosh(\frac{x\tau}{\cos\theta})(x\cos x-\sin x)}
{(\sinh(\frac{x\tau}{\cos\theta})\frac{\cos\theta}{\tau})^3}
\nonumber \\
&&\langle\sin kJ_1(\frac{V_b\sin x\sin k}{\cos\theta})
J_0(\frac{V_b\sin x\sin k}{\cos\theta})\rangle_k
f_z(\frac{x}{\cos\theta}),
\label{Mz}
\end{eqnarray}
Unlike $\Delta C$, $\Delta M_z(\theta)$ has a limit of
$T=0$ which is shown in Fig.~3 at $V_b=20$ and $V_c=2$.  Note
that the magnetic energy $H_z\Delta M_z$ may be much greater than the
Zeeman energy $(\mu_BH)^2/\varepsilon_F$ and the orbital energy
$(t_b^2\cos^2\theta+t_c^2\sin^2\theta)\Omega^2/12\varepsilon^3_F$ of the
non-interacting electrons (the latter formula is a result beyond
quasi-classics).  The magnetization is found to be negative
(diamagnetic) in agreement with the experiments in the high field
reentrant metallic phase of (TMTSF)$_2$ClO$_4$ \cite{YC}.  The torque is
determined by the expression $N_x=-\Delta M_zH\cos\theta+\Delta
M_yH\sin\theta$, where $\Delta M_y$ is given by Eq. (\ref{Mz}) after
interchanging $t_b$ with $t_c$ and $\cos\theta$ with $\sin\theta$.  The
torque is plotted in Fig. 3 at $\tau=0$ and $\tau=0.25$ together with an
experimental curve taken at $T=0.4$ K, $H=5$ T \cite{ND}.  Using the
cited above value of $\Omega/H$, one finds $\tau=0.28$ in the
experiment, close to the value employed for the theoretical curve c.
There is an overall qualitative resemblance between the theoretical and
the experimental curves.  For a detailed comparison one should take into
account the different spacings of the chains along the $y$- and $z$-axes
and the triclinic structure of the real material.  Note that, with few
exceptions, the peaks in Fig. 3 appear when $\tan\theta$ is integer, not
fractional as in Fig. 2, in agreement with the experimental observation
\cite{N}.  This happens because at $\tau=0$ the aperiodic function of
$x$ under the integral (\ref{Mz}) decreases rapidly when $x\geq 2\pi$,
while the last function in (\ref{Mz}) has the (longer) period of the
order of $\pi/\tan\theta$, thus the resonances cannot take place at
small $\tan\theta$.

   In Fig. 4, $\Delta M_z$ (\ref{Mz}) at $\theta=0$ is plotted vs the
variable $1/V_b$, which is proportional to the magnetic field. $\Delta
M_z$ exhibits oscillations periodic in inverse magnetic field, which may
correspond to the fast oscillations of magnetization observed in Ref.
\cite{YC}.  The period of the oscillations $\Delta(1/H)$ is determined
by the condition $\Delta V_b=\pi$ and equals $\pi ebv_F/8ct_b$.  The
numbers of oscillations in Fig. 4 are from 3 to 15.  Using the
experimental value 255T for the frequency of the fast oscillations
\cite{YC} one finds $t_b=180$K.  A similar mechanism for the fast
oscillations of conductivity was studied in Ref. \cite{Ya}.  However, we
have not found these oscillations in specific heat (\ref{DC}).

   Another possible explanation of the fast oscillations \cite{BY} takes
into account that the unit cell of (TMTSF)$_2$ClO$_4$ is doubled in the
$y$ direction due to a crystal superstructure, thus the energy of a
chain staggers by the value $\pm\kappa$ along the $y$-axis.  Suppose,
that $\kappa\gg t_b$.  In this case, the conduction
band is split into two bands with the Fermi momenta $k_F\pm\kappa/v_F$,
and the effective transverse hopping integral $t_b^*$ is equal to
$t_b^2/2\kappa$.  In one band, the electron wave functions are
predominant on even chains; in another band, on odd chains.  Interaction
between the electrons belonging to the same band reproduces the formulas
given above with $t_b^*$ substituted for $t_b$ and the spacing along
$y$-axis doubled.  This prescription immediately explains why in
experiments \cite{N,Oe} the angular oscillations were found only at
even, instead of all, integer values of $\tan\theta$ defined with
respect to the original lattice without superstructure.  In addition,
the interaction of the electrons belonging to the different bands gives
the following contribution to the free energy (at $\theta=0$): \FL
\begin{eqnarray}
&& \Delta'F=-\frac{2\pi^3g^2T^3t_b^4}{\varepsilon_F\Omega\kappa^4}
\int_0^\infty dx\frac{\cosh(x\tau')}{\sinh^3(x\tau')}
\langle J_0^2(V_cx\sin p)\rangle_p \nonumber \\
&&\langle[(1+2\cos^2x)J_0^2(V_b^*\sin x\sin k)
-J_1^2(V_b^*\sin x\sin k)] \nonumber \\
&&\sin^2k\cos(2Rx)+4J_0(V_b^*\sin x\sin k)
J_1(V_b^*\sin x\sin k) \nonumber \\
&&\sin k\cos x\sin(2Rx)\rangle_k ,
\label{ka}
\end{eqnarray}
where $V_b^*=4t^*_b/\Omega$, $\tau'=\tau/2$ and $R=2\kappa/\Omega$.
The correction to the specific heat $\Delta'C$ is obtained from
(\ref{ka}) in the same way as Eq. (\ref{DC}) is obtained from Eq.
(\ref{DF}) and is plotted vs $R$ in Fig. 5. $\Delta'C$ exhibits
oscillations periodic in inverse magnetic field which may correspond to
the fast oscillations of specific heat found in Ref. \cite{F}.  The
period of the oscillations is determined by the condition $\Delta R=1$
and is given by $\Delta(1/H)=ebv_F/2c\kappa$, which corrects by a factor
of 2 the expression found in Ref. \cite{BY}.  The numbers of
oscillations in Fig. 5 are from 2 to 40.  Using the experimental
frequency of the fast oscillations 257T \cite{F} one finds
$2\kappa=460$K.  Analogous oscillations should also exist in
magnetization, which can be derived from Eq. (\ref{ka}) as well.  If, on
the other hand, the opposite limit $\kappa\leq t_b$ holds, then, in
order to explain why the odd peaks are missing in experiments
\cite{N,Oe}, it is necessary to conclude that the field $H_z$ at $H=5$ T
is below the magnetic breakdown field $H_0=\kappa^2c/t_bebv_F$.  In this
case, the mechanism described above still works qualitatively but with
the frequency of the fast oscillations proportional to $t_b$ instead of
$\kappa$.  If the odd peaks appear experimentally in higher fields, this
can give the estimates of $H_0$ and $\kappa$.

   Fast oscillations were also observed in the material
(TMTSF)$_2$PF$_6$, which does not have an anion superstructure
\cite{PF6}.  This fact supports the first explanation of the fast
oscillations which does not invoke the superstructure.  On the other
hand, these oscillations were observed thus far only in the FISDW phase
in transport measurements.  Thus, formally, the fast oscillations in
(TMTSF)$_2$PF$_6$ are outside the applicability range of presented
theory and a definitive choice between two explanations cannot be made
(see also Ref. \cite{L2}).

   In conclusion, the second order correction to the free energy due to
the interaction between electrons has been calculated for a
quasi-one-dimensional metal.  Angular and fast magnetic oscillations of
specific heat, magnetization and torque were found in qualitative
agreement with experiment.  Two different mechanisms of the fast
oscillations, related and unrelated to the specific crystal structure of
(TMTSF)$_2$ClO$_4$, were observed.  Unlike in the standard theory of
metals, the oscillations are completely due to the interactions between
electrons and cannot be interpreted in terms of the presence of some
closed electron orbits.  The oscillations involve the energy scales of
the cyclotron frequency $\Omega$, the hopping integral $t_b$ or the
anion superstructure splitting $\kappa$, which are much higher than the
characteristic energy scale of the deviation from nesting and the FISDW
transition temperature.  That explains why the oscillations persist in
the FISDW phase as observed experimentally.

   The author thanks M. J. Naughton for sending torque data shown in
Fig. 3; J. S. Brooks, who suggested to calculate specific heat; both of
them and P. M. Chaikin, for discussions.  This work was supported by the
NSF Grant No. DMR 89-06958.

\figure{ The second order correction to the free energy.  The lines,
labeled by + and --, represent Green functions of the electrons with
momenta close to $\pm k_F$. The vertices represent the amplitudes of
interaction between electrons. \label{fig1}}

\figure{ Normalized correction to specific heat $-\Delta C/3g^2C_0$ vs
magnetic field rotation angle $\theta$. Parameter
$\tau$ is equal to 0.1 (curve a), 0.2 (b) and 0.4 (c). The arrows at
the top indicate the rational values of $\tan\theta$. \label{fig2}}

   \figure{ Normalized magnetization $-16\Delta
M_zH\varepsilon_F/g^2\Omega^2$ (curve a) and normalized torque
$16N_x\varepsilon_F/g^2\Omega^2$ (curve b, $\tau=0$, and curve c,
$\tau=0.25$) calculated as functions of $\theta$.  The curve d
represents the experimental torque data due to M. J. Naughton \cite{ND}.
\label{fig3}}

   \figure{ Normalized magnetization $-16\Delta
M_zH\varepsilon_F/g^2\Omega^2$ at $T=0$ vs normalized magnetic field
$1/V_b=ebv_FH/8t_bc$; $t_c=0$ in curve a and $t_c=0.1t_b$ in curve b.
\label{fig4}}

   \figure{ Additional normalized correction to specific heat in the
presence of the special crystal superstructure
$-\Delta'C\kappa^4/6g^2t^4C_0$ vs normalized inverse magnetic field
$R=2\kappa c/ebv_FH$.  Parameters are: $t_b^*=\kappa/8,\;t_c=\kappa/80$
and $T=\kappa/20\pi^2$. \label{fig5}}


\begin{references}

\bibitem[*]{*} On leave from L. D. Landau Institute for
Theoretical Physics, Moscow, 117940, Russia.

   \bibitem[1]{N} M. J. Naughton et al., Phys.  Rev. Lett. {\bf 67},
3712 (1991).

   \bibitem[2]{YC} X. Yan et al., Phys.  Rev. B {\bf 36}, 1799 (1987);
R. V. Chamberlin et al., Phys.  Rev. Lett. {\bf 60}, 1189 (1988).

   \bibitem[3]{F} N. A. Fortune et al., Phys.  Rev. Lett. {\bf 64},
2054 (1990).

   \bibitem[4]{LAK} I. M. Lifshits, M. Ya. Azbel and M. I. Kaganov, {\it
Electron Theory of Metals} (Consultants Bureau, N.Y., 1973).

   \bibitem[5]{L} A. G. Lebed, Pis'ma Zh. Exp. Teor.  Fiz. {\bf 43},
137 (1986) [JETP Lett. {\bf 43}, 174 (1986)].

   \bibitem[6]{XY} X. Yan et al., Solid State Commun. {\bf 66}, 905
(1988).

   \bibitem[7]{LB} A. Lebed and P. Bak, Phys.  Rev. Lett. {\bf 63},
1315 (1989).

   \bibitem[8]{MRS} G. S. Boebinger et al., Phys.  Rev. Lett. {\bf 64},
591 (1990); M. J. Naughton et al., Mater.  Res. Soc. Symp.  Proc. {\bf
173}, 257 (1990).

   \bibitem[9]{Oe} T. Osada et al., Phys.  Rev. Lett. {\bf 66}, 1525
(1991).

   \bibitem[10]{Ot} T. Osada, S. Kagoshima and N. Miura, Technical
report of ISSP, Univ. of Tokyo, Ser. A, No. 2470 (1991); K. Maki, Phys.
Rev. B {\bf 45}, 5111 (1992).

   \bibitem[11]{M2} A. Bjeli\u{c} and K. Maki, Phys. Rev. B, to be
published.

   \bibitem[12]{GL} L. P. Gor'kov and A. G. Lebed', J. Phys. (Paris)
Lett. {\bf 45}, L433 (1984).

   \bibitem[13]{Y2} V. M. Yakovenko, Zh. Exp. Teor.  Fiz. {\bf 93}, 627
(1987) [Sov.  Phys.  JETP {\bf 66}, 355 (1987)].

   \bibitem[14]{ND} The author is grateful to M. J. Naughton for
permission to publish the raw experimental torque data of Ref.
\cite{N}.  The curve d in Fig. 3 actually represents the data between
two main minima in the angle interval from $6^\circ$ to $-90^\circ$
mapped for the purpose of qualitative comparison to the interval
$0^\circ$ to $90^\circ$.

   \bibitem[15]{Ya} K. Yamaji, J. Phys.  Soc. Jpn. {\bf 55}, 1424
(1986).

   \bibitem[16]{BY} S. A. Brazovskii and V. M. Yakovenko, Pis'ma Zh.
Exp. Teor.  Fiz. {\bf 43}, 102 (1986) [JETP Lett. {\bf 43}, 134 (1986)].

   \bibitem[17]{PF6} J. P. Ulmet et al., J. Phys. (Paris) Lett. {\bf
46}, L535 (1985); J. R. Cooper et al., {\em Phys.  Rev. Lett.} {\bf63},
1984 (1988); S. T. Hannahs et al., ibid. {\bf63}, 1988 (1988);
W. Kang et al., Phys. Rev. B, to be published.

\bibitem[18]{L2} A. G. Lebed', Phys. Scripta {\bf 39}, 386 (1991).

\end{references}
\end{document}